\documentclass{aa}
\usepackage{txfonts}
\usepackage{amssymb}
\usepackage{natbib}
\bibpunct{(}{)}{;}{a}{}{,}
\usepackage[dvips]{graphicx}
\usepackage[dvips]{graphics}
\usepackage[english]{babel}
\usepackage[latin1]{inputenc}
\usepackage{subfigure}
\usepackage{longtable,lscape}
\usepackage{supertabular}
\usepackage{appendix}
\newcommand{\excs}{\extracolsep{\fill}}

\title{MESS (Multi-purpose Exoplanet Simulation System)\thanks{\bf www.messthecode.com} }
\subtitle{A Monte Carlo tool for the statistical analysis \\
and prediction of exoplanets search 
results.\thanks{send offprint requests to bonavita@astro.utoronto.ca}}
\author{M. Bonavita \inst{1,2}, G. Chauvin \inst{3,4}, S. Desidera \inst{1}, 
R.Gratton \inst{1}, M. Janson \inst{2}, J.~L. Beuzit \inst{3}, M. Kasper \inst{5}, C. Mordasini \inst{4}}
\institute{INAF - Osservatorio Astronomico di Padova, Vicolo 
dell'Osservatorio 5, 35122 Padova, Italy 
\and Department of Astronomy and Astrophysics - University of Toronto, 50 St. George Street M5S 3H4 Toronto ON Canada 
\and UJF-Grenoble 1 / CNRS-INSU, Institut de Planetologie et d'Astrophysique de Grenoble (IPAG) UMR 5274, Grenoble, F-38041, France
\and Max Planck Institute for Astronomy, K\"onigstuhl 17, D-69117 Heidelberg, Germany
\and European Southern Observatory (ESO), Karl-Schwarzschild-Str. 2, 85748 Garching, Germany}

\date{Received/Accepted}
\begin{document}
\abstract
{The high number of planet discoveries made in the last years provides
a good sample for statistical analysis, leading to some clues on the 
distributions of planet parameters, like masses and periods, at least 
in close proximity to the host star. We likely need to wait for 
the extremely large telescopes (ELTs) to have an 
overall view of the extrasolar planetary systems. Those facilities will finally 
ensure an overlap of the discovery space of direct and indirect techniques, 
which is desirable to completely understand the nature of the discovered 
objects, obtaining both orbital parameters and physical characterization.}
{In this context it would be useful to have a tool that can  be used for 
the interpretation of the present results obtained with various observing techniques,
 and also to predict what the outcomes would 
be of the future instruments.} 
{For this reason we built MESS: a Monte Carlo simulation code which uses 
either the results of the statistical analysis of the properties of 
discovered planets, or the results of the planet formation theories, 
to build synthetic planet populations
fully described in terms of frequency, orbital elements and physical
properties. 
They can then be used to either test the consistency of
their properties with the observed population of planets given
different detection techniques (radial velocity, imaging and
astrometry) or to actually predict the expected number of planets for
future surveys, as well as to optimize the future multi-techniques
observations for their characterization down to telluric masses.}
{In addition to the code description, we present here some
of its applications to actually probe the physical and orbital
properties of a putative companion within the circumstellar disk of a
given star and to test constrain the orbital distribution properties
of a potential planet population around the members of the TW Hydrae
association. Finally, using in its predictive mode, the synergy of
future space and ground-based telescopes instrumentation has been
investigated to identify the mass-period parameter space that will be
probed in future surveys for giant and rocky planets}
{}

\titlerunning{MESS (Multi-purpose Exoplanet Simulation System)}
\authorrunning{Bonavita et al.} 

\maketitle

\section{Introduction}
\label{sec:intro}

Many statistical studies have been done using information coming 
from more than a decade of extensive searches for exoplanets, trying 
to answer questions either related to the properties of those objects, 
such as the mass, orbital period and eccentricity 
\citep{2003ApJ...598.1350L,2008PASP..120..531C}, or 
about the relevance of the host star characteristics (mass, 
metallicity and binarity) on the final frequency and distribution of planetary 
systems \citep[see][]{2005ApJ...622.1102F, 2004A&A...415.1153S, 2007ApJ...670..833J}.
Since the most successful techniques (radial velocity and transit) 
have focused on the inner ($\le 5~AU$) environment of main sequence 
solar-type stars, most of the available information on the frequency 
of planets concern this class of stars. 

Recent discoveries of young distant planetary mass objects with 
direct imaging \citep[see e. g.][]{2008Sci...322.1348M, 2008Sci...322.1345K,2008arXiv0811.3583L} 
are giving us a first hint on the potential of the direct detections 
in the exploration of the outer region of the planetary systems, also 
raising many questions about how such objects could form \citep[see][]{2009A&ARv.tmp...16A}.  
This defines the niche of the next generation high contrast imaging 
instruments like the Gemini Planet Imager \citep[GPI:][]{2007AAS...211.3005M}
and VLT/SPHERE \citep[Spectro-Polarimetric High-contrast Exoplanet 
REsearch:][]{2008SPIE.7014E..41B}. These instruments will likely allow us 
to extend such a systematic characterization to larger scales ($\ge 10~AU$). 
Due to practical limitations (inner working angle, best contrast achievable), 
these instrument will focus on warm giant planets on orbits far away from 
their stars, paving the path for the ELTs facilities. A wide range of 
planetary masses and separations, down to the rocky planets (and, in very 
favourable cases reaching the habitable zone), will be explored with 30-40 
meter-class telescopes, finally allowing an overlap between the discovery 
spaces of direct and indirect techniques.

In this context it is useful and crucial to predict the performances of 
the forthcoming instruments, not only in terms of number of expected 
detections, but also trying to figure out what will be the explored 
parameter space and even the possible synergies between different discovery 
techniques. 

Here we present our Monte Carlo simulation code MESS, whose aim is to provide 
a flexible and reliable tool for the statistical analysis and prediction
of the results of planet searches.

It produces synthetic planet populations, deriving all the physical parameters
of these planets together with the observables that can be compared with 
the predicted capabilities of existing or planned instruments. Such comparisons
allow to derive subsets of fully characterized {\it detectable planets}, as
well as a snapshot of what the evolution of the sample of detected 
planets would be in the next years.

A detailed description of the code, and of all the assumptions which 
constitute its basis, is given in Sect.~\ref{sec:mess}, while in 
Sect.~\ref{sec:op_modes} we present the different operation modes of the code and their
applications. 
 Although the MESS has been built, and it has been so far applied, only to 
analyze and/or predict the results of direct imaging surveys, an extension
of the code to different techniques is planned. 
The first attempt in this direction are presented in Sec.~\ref{sec:pred_om}.
Conclusions and
suggestions for further work will be finally drawn in Sect.~\ref{sec:conclusions}.

\section{MESS (Multi-purpose Exoplanet Simulation System)}
\label{sec:mess}

Over the past years, several groups \citep{2007A&A...472..321K,2008ApJ...689L.153L,2010A&A...509A..52C,2010ApJ...717..878N} 
initiated statistical analysis to constrain the physical and orbital 
properties (mass, period, eccentricity distributions) of the giant planet 
population at large separations. They developed statistical analysis tools 
appropriate to exploit the performances of deep imaging surveys. They also 
tested the consistency of various sets of parametric distributions of 
planet properties, using the specific case of a null detection. The first 
assumption of these tools is that planet mass and period 
distributions coming from the statistical results of RV studies at short 
period \citep[see e.g.][]{2003ApJ...598.1350L,2008PASP..120..531C} can be 
extrapolated and normalized to obtain information on more distant planets.
Despite the model-dependency on the mass predictions, the approach is attractive
for exploiting the complete set of detection performances of the survey
and characterizing the outer portions of exo-planetary systems.

With all of this in mind, we tried to go a step further, creating a 
Multi-purpose Exo-planet Simulation System (hereafter MESS) to be applied also to other techniques than direct imaging, also using the information coming from the planetary formation theories. 

The code is written in IDL and can be downloaded from www.messthecode.com

The basics operations performed by the code are the following: 

\begin{enumerate}
\item it generates a synthetic population of planets, including all the orbital elements, 
either using the planet mass and period distributions coming from the 
statistical results of RV studies or the outcome of the planetary formation theories.
\item taking into account the characteristics of the host star and of the planetary orbit, it calculates all the observable quantities needed for the comparison with the instrument performances, such as radial velocity (RV) and astrometric signal, planet/star contrast, degree of polarization, etc.
\item given the detection capability relation of an instrument, either already available or planned, it selects a sub-sample of fully characterized {\it detectable planets}, which characteristics can then be analyzed. 
\end{enumerate}

\noindent The code then assumes a given star population, a planet population
with associated physical and orbital properties based on a theoretical
or semi-empirical approach, the corresponding observables for
different observing techniques, finally generate a synthetic population
of planets to be compared with the instrumental detection
performances. Each step is described hereafter.

\subsection{Star population}
\label{sec:star_pop}

The first input of the MESS is a sample of $N_{\rm Star}$ stars, 
which have been targeted for planet searches or 
which are part of a sample for future observations.
Various stellar parameters are assumed to
be known, such as the apparent magnitude, the distance, the luminosity, the spectral type, the mass, the 
age, the metallicity, etc.

\begin{figure}[h]
\includegraphics[width=9.2cm]{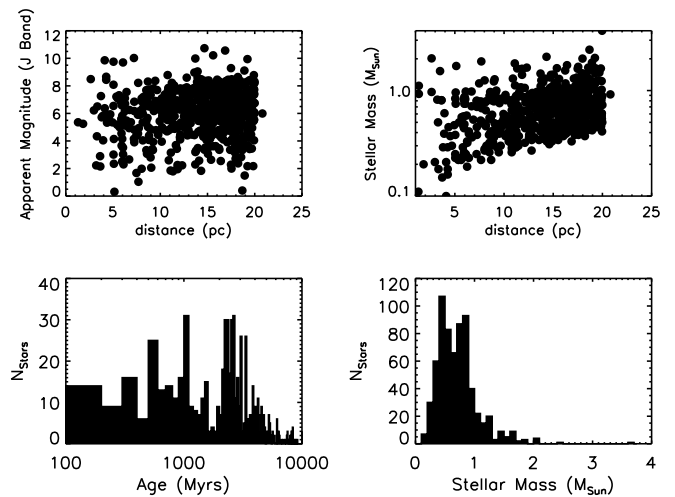}
\caption{\footnotesize Principal characteristics of the sample of nearby 
stars used to built the example synthetic population. {\bf Upper Left:} Apparent 
Magnitude in the J Band vs distance in pc. {\bf Upper Right:} Stellar Mass 
($M_{\odot}$) vs distance in pc. {\bf Lower Left:} Histogram of stellar ages 
(Myrs). {\bf Lower Right:} Histogram of stellar masses ($M_{\odot}$).}
\label{fig:sample}
\end{figure}

Fig.~\ref{fig:sample} shows the characteristics of a sample of 600 nearby ($d < 20$ pc)
stars selected from the Hipparcos catalogue \citep{1997ESASP1200.....P} and used to build the synthetic population
showed in Fig.~\ref{fig:pop_char_I}.

\subsubsection{Binarity module}
MESS also gives the possibility of taking into account the presence of one (or more) 
additional stellar companions, in the analysis.
If a star in the sample is flagged as binary, the code uses the information about the binary 
orbit (if available) to compute the critical semi-major axis for the dynamical stability of the system.
This set the limiting value that the semi-major axis of a planet
can attain and still maintain its orbital stability, as a function of the mass-ratio 
and orbital elements of the binary, as shown by \cite{1999AJ....117..621H}.

Both the case of circumstellar (or Satellite S-type) and circumbinary (or Planet P-type) orbit are considered, and the critical semi-major axis is computed
using Eq.~\ref{eqn:acrit_cs} and \ref{eqn:acrit_cb} respectively, from \cite{1999AJ....117..621H}.

\begin{eqnarray}
 \label{eqn:acrit_cs}
 {a_c}/{a_b} & = & 0.464 - 0.38  ~ \mu - 0.361  ~ e_b +  0.586  ~ \mu  ~ e_b \nonumber \\
 & + & 0.150  ~ \mu^2 -  0.198  ~ \mu  ~e_b^2 
\end{eqnarray}

\begin{eqnarray}
\label{eqn:acrit_cb} 
{a_c}/{a_b} & = & 1.60 + 4.12 ~\mu + 5.10~e_b - 4.27 ~\mu ~e_b + \nonumber \\
 & - & 5.09 ~\mu^2 - 2.22 ~e_b^2 + 4.61 e_b^2~\mu^2 
\end{eqnarray}

\noindent
In both the equations, $a_c$ is {\it critical} semi-major axis , 
$\mu={{M_1}/{({M_1}+{M_2}})}$, 
$a_b$ and $e_b$ are the semi-major axis and eccentricity of the binary, and 
$M_1$ and $M_2$ are the masses of the primary and secondary stars,
respectively. 
If not available from literature, the eccentricity is assumed to be $e_b=0.36$, 
reported as mean value for the eccentricity of a 
binary system, by \cite{1991A&A...248..485D}.
In case the value of the semi-major axis is not available, then the code estimates it 
as $a_b=1.31~\rho~(arcsec)~d~(pc)$ \citep[see][]{2002PASP..114..529F,1991A&A...248..485D}\footnote{Note that \cite{1991A&A...248..485D} refers to solar-type star multiplicity}.

Note that in the first case (S-type orbit), $a_c$ set the  maximum
value that the semi-major axis of a planet can assume, before compromising the stability,
while it represents the minimum value of the semi-major axis of a stable planet, in the case of a P-type orbit.

\subsection{Planet population}
\label{sec:ppop}
The core of the code is the generation of the synthetic planets, that are fully characterized, both in terms
of orbital parameter, and physical characteristics.
Depending on the goal of the study one can choose between a {\it Semi-empirical approach} or a {\it Theoretical Approach}.
These different approaches makes the code suitable to constraint the planet properties under different assumptions, but also to test model predictions. 

\noindent If the {\it Theoretical Approach} is chosen, masses and period values selected 
from a synthetic population provided by the 
output of the planetary formation models \citep[see e.g. ][]{2009A&A...501.1139M} are given as input.
In this case all the orbital characteristics are also provided, together with the 
physical properties of each planet, so no random generation is needed, and the code only 
evaluates the observable and compares them to the provided detectability relations. 
Different populations of planets obtained assuming different stellar 
masses and metallicity values can be selected according with the characteristics 
of the real star in the sample, to take into account the effects of the 
stellar characteristics on the planet formation 
processes.\footnote{The results discussed in this paper has been obtained using mainly the Semi-empirical Approach. 
An extensive use of the Theoretical Approach, using as input the newest outcomes of the Bern formation models (Mordasini et al. 2011), will be the subject of a forthcoming paper.}.

The {\it Semi-empirical Approach} uses the power law distributions in Eq.~\ref{eqn:powlm} and ~\ref{eqn:powlp} for the mass and semi-major axis of the planets
as retrieved from the statistical analysis of the properties
of the planets discovered so far to generate a seed population of $N_{seed}$ values of masses and periods (see Sec.~\ref{sec:mpseed}).

\begin{center}
\begin{eqnarray} 
\label{eqn:powlm}
\frac{dN}{d\left(M_{\rm p}\right)} & \propto & {\left(M_{\rm p}\right)^{\alpha}} \\
\frac{dN}{dP}
                              & \propto & P^{\beta}
\label{eqn:powlp}
\end{eqnarray} 
\end{center}

The user can also set a pre-determined grid of masses-periods and feed it to the code, 
without any assumption on the distributions. 
This would be the case if, for example, boundaries on the mass/semi-major axis space where planets can form 
are to be set using the outcomes of a formation model\citep[see e.g][]{2010arXiv1012.5281M}, 
excluding from the sample the planets not compatible with the theory. 

If the Semi-empirical approach is used, mass, orbital
parameters, as well as temperature and radius of the planets, are
obtained based on the assumption described in the next sections

\subsubsection{Mass-period seed generation}
\label{sec:mpseed}
If the semi-empirical approach is chosen, 
the power law distributions are fed to the Monte-Carlo core of the code, 
that randomly generates a fixed number of mass-period pairs.
Both the planetary mass and period ranges can be given as inputs, 
together with the power-law exponents. In a typical setup, the power-law 
exponents are assumed to be $\alpha=-1.31$ and $\beta=-0.74$ respectively, 
from \cite{2008PASP..120..531C}. The planetary masses span the range 
between 0.6~$M_{Earth}$ and 15~$M_{Jup}$, and the periods ($P$) are chosen 
between 2.5 days and 350 years (corresponding to 50 AU for $1~M_{\odot}$ star).

A scaling of the planetary mass, and even of the period, with the stellar 
mass can be also introduced, according to recent results  \citep[e.g.][]{2007A&A...472..657L}. 
In addition, a dependence of the planet frequency on the stellar metallicity 
may also be considered \citep[see][]{2005ApJ...622.1102F}.

\subsubsection{Evaluation of the orbital parameters}
\label{sec:orb_param}

For each mass-period pair in the seed generation, the code 
evaluates the semi-major axis computed using Kepler's 
third law, using the mass of each star in the input sample.
Then, it generates $N_{gen}$ values of all the
orbital parameters: eccentricity ($e$), inclination ($i$), longitude of periastron
($\omega$), longitude of ascending node ($\Omega$), and time of periastron passage ($T_0$).
By default, all these parameters are randomly generated following an uniform 
distribution \footnote{Note that in the case of the inclination, 
$\cos i$ and not $i$ itself is uniformly generated by the code}.
The eccentricity distribution is cut at $e=0.6$ as suggested
by the results of the RV surveys (see Cumming et al. 2008). This also allows 
to control possible bias towards high eccentricity planets that could affect the 
results of Direct Imaging surveys. A full discussion of the impact of the eccentricity distribution on the simulations results is held in Sec.~3.4.

The date of observation is also required. 
If not available from the real data, an epoch of observation, $t_{obs}$, is 
generated over a time-span chosen according with the considered instrument. 

The code also offers the possibility to fix each orbital parameter to known 
or predicted values, for all the planets in the population.

The coordinates, $x$ and $y$, of the projected orbit on the plane 
perpendicular to the line of sight, are finally computed using the ephemeris formulae 
of \cite{1978GAM....15.....H}, reported in  Eq.~\ref{eqn:xy_proj} to \ref{eqn:rho_proj}.

\begin{eqnarray}
x & = & AX+FY       \label{eqn:xy_proj}\\
y & = & BX+GY \nonumber \\
X & = & \cos E -e    \label{eqn:xy_true} \\
Y & = & \sqrt{1-e^2}\sin E \nonumber \\
\rho & = & \sqrt{ x^2 + y^2} \label{eqn:rho_proj}
\end{eqnarray}

\noindent where $X$ and $Y$ are the coordinates of the orbit (Eq. \ref{eqn:xy_true}),
$\rho$\ is the projected separation,  and $A, B, F, G$ are the Thiele-Innes elements, 
which can be obtained from the classical ones (the semi-major axis $a$, $\omega$, 
$\Omega$, and $i$) using Eq.~\ref{eqn:abfg}:
\begin{eqnarray}
A & = & a\left(\cos \omega \cos \Omega - \sin \omega \sin \Omega \cos i\right) \nonumber\\
B & = & a\left(\cos \omega \sin \Omega + \sin \omega \cos \Omega \cos i \right) \label{eqn:abfg}\\
F & = & a\left(-\sin \omega \cos \Omega - \cos \omega \sin \Omega \cos i \right)\nonumber \\
G & = & a\left(-\sin \omega \sin \Omega + \cos \omega \cos \Omega \cos i \right)\nonumber.
\end{eqnarray}
\noindent In these equations, $E$\ is the eccentric anomaly (obtained from the mean 
anomaly $M$\ ( Eq. \ref{eqn:m_anom}) using Eq. \ref{eqn:ecc_an}) and $\nu$\ the true 
anomaly (Eq. \ref{eqn:tr_anom}):
\begin{eqnarray}
M   & = & \left(\frac{t_{obs}-T_0}{p}\right)2\pi \label{eqn:m_anom}\\
E_0 & = & M + e \sin M + \frac{e^2}{2} \sin 2M \nonumber \\
M_0 & = & E_0 - e\sin E_0 \nonumber\\
E   & = & E_0 + \left(M-M_0)\right)/\left(1-e\cos E_0\right)\label{eqn:ecc_an}\\
\tan{\nu/2} & = & \sqrt{ \left(1+e\right)/\left(1-e\right)} \tan{E/2} \label{eqn:tr_anom}
\end{eqnarray}

The projected separation, $\rho$ (in arcsec), can be obtained either using 
Eq.~\ref{eqn:rho_proj} or Eq. \ref{eqn:rho_r} (which gives also an estimate of 
the radius vector: $r$), then dividing for the star distance.
\begin{eqnarray}
\rho & = & r \cos \left(\nu + \omega\right) \sec \left( \theta - \Omega\right)  \label{eqn:rho_r}\\
r    & = & a \left( 1-e^2 \right)/\left(1+e \cos \nu \right) \nonumber
\end{eqnarray}

\subsubsection{Planet Temperature}

Since we aim at consider both the thermal and reflected flux of the planets, 
we need two different estimates of the temperature. The first one is the 
internal temperature, $T_{\rm int}$, coming from the evolutionary models 
\citep[see e.g.][]{2003A&A...402..701B}. The second
one is the equilibrium temperature, $T_{\rm eq}$, obtained through 
Eq. \ref{eqn:t_eq} \citep[from][]{2003ApJ...588.1121S}
\begin{equation}
\label{eqn:t_eq}
T_{\rm eq}=\left[ \frac{ \left(1-A_{\rm B} \right) L_*} {16 \pi \sigma a^2} \right],
\end{equation}
\noindent 
where $L_*$\ is the star luminosity. The Bond albedo $A_{\rm B}$\ is assumed to 
be 0.35 in the J band \citep[Jupiter value, see][]{1981JGR....86.8705H} and it is randomly generated 
between 0.3 and 0.52 in the visible \citep[the latter being the Jupiter albedo 
in V band,see][]{2003ApJ...588.1121S}

Our final assumed value for the effective temperature of the planet $T_{\rm eff}$\
is given by:
\begin{equation}
\label{eqn:t_pl}
T_{\rm eff}^4 = T_{\rm int}^4 + T_{\rm eq}^4.
\end{equation}

\subsubsection{Planet radius}

\begin{figure}[tbp]
\center
\includegraphics[width=9.2cm]{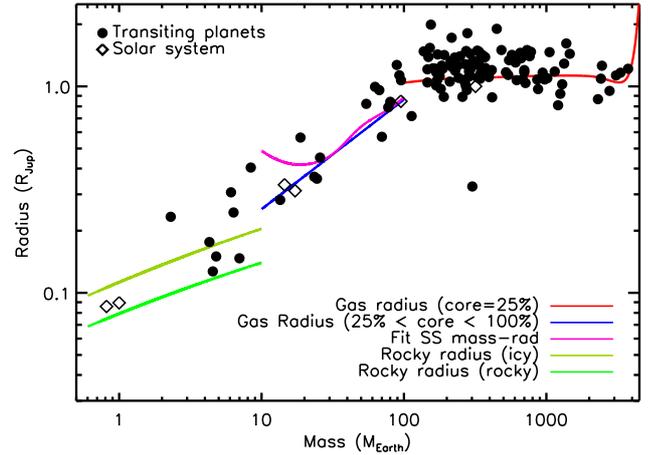}
\caption{\footnotesize Summary of the planetary Mass - Radius relations adopted 
for the different mass ranges. All the model computation are made assuming a 
host star of 1~$M_{\odot}$, and the semi-major axis value is fixed to 5~AU.
Filled symbols corresponds to known transiting planets;
 open symbols are for Solar System planets.}
\label{fig:mr_rel}
\end{figure}

MESS uses the approach developed by \cite{2007ApJ...659.1661F} to evaluate the 
planetary radius. Practically the radius is assumed to depend on the planet mass,
with the following recipes:

\begin{enumerate}
\item For Jupiter-like planets ($M \ge 100 M_{\rm Earth}$ ), an interpolation is 
performed within the published values given by \cite{2007ApJ...659.1661F}. Values 
of age and distance of each star are entered, yielding a value for $R_{\rm Gas}$. 
A core mass of $10 M_{\rm Earth}$ is assumed.

\item Equations \ref{fit1} and \ref{fit2} from  \cite{2007ApJ...659.1661F} are used
for the smallest planets ($M \le 10 M_{\rm Earth}$). These are either:
\begin{eqnarray}
\label{fit1}
R & = & (0.0912 \quad{imf} + 0.1603)(\log{M})^2 \nonumber\\
  &   & +(0.3330 \quad{imf} + 0.7387)\log{M} \nonumber\\ 
  &   & +(0.4639 \quad{imf} + 1.1193) \quad \quad 
\end{eqnarray}
or:
\begin{eqnarray}
\label{fit2}
R & = & (0.0592 \quad{rmf} + 0.0975)(\log{M})^2 \nonumber\\
  &   & +(0.2337 \quad{rmf} + 0.4938)\log{M} \nonumber\\ 
  &   & +(0.3102 \quad{rmf} + 0.7932) \quad \quad 
\end{eqnarray}
for ice/rock and rock/iron planets, respectively. In these equations, $R$\ is 
in $R_{\rm Earth}$ and $M$ is in $M_{\rm Earth}$, while ${imf}$ is the ice mass 
fraction (1.0 for pure ice and 0.0 for pure rock) and ${rmf}$ is the rock mass 
fraction (1.0 for pure rock and 0.0 for pure iron). In the typical MESS setup, 
the ice/rocky or rocky/iron fraction is set to 0.3 (50\% of chance for a planet 
being mainly icy or rocky). 
\item Finally, predictions are uncertain for the Neptune-like planets, where 
the transition between the two relations described above should occur. The most 
sensible approach seems to be to fit the mass-radius relation of the Solar System in 
the same mass-range ($10-40M_{\rm Earth}$ ). This procedure provides a good 
agreement with the radii of Uranus and Neptune and of the few transiting 
Neptunes confirmed so far (as listed by The Extrasolar
Planet Encyclopaedia\footnote{www.exoplanet.eu} see Fig. \ref{fig:mr_rel}).
\end{enumerate}

The resulting mass-radius relations are showed in Fig. \ref{fig:mr_rel}, with 
over-plotted the data corresponding to the planets discovered with the transit 
technique and the planets from our Solar System, for comparison.

\subsection{Predicted observables}
\label{pla_char}
Having in hands the full set of orbital and physical parameters of the planets, 
the code then provides an estimate of observable quantities such as the luminosity 
contrast or the degree of polarization, needed for direct observations, but also quantifies 
the  indirect effects of the presence of the planet, 
providing a measure of the semi-amplitude of radial velocity (RV) and the astrometric signal. 
 
\subsubsection{Planet/Star contrast}
\label{sec:contrast}
MESS gives an estimate of both the intrinsic and reflected flux, 
in the selected band, for each planet. 
Throughout the paper we will refer to the planets which luminosity is dominated by the 
intrinsic contribution as self-luminous or {\it warm} planets, as opposed to the {\it cold} 
planets for which the reflected light provides most of the contribution to the planet/star contrast.

The intrinsic emission is estimated using the prediction of evolutionary 
models at the age of the star (assumed to be also the age of the system). To 
this purpose two classes of models can be considered, based on different 
assumptions on the initial conditions: Hot Start models 
\citep{2000ApJ...542..464C,2003A&A...402..701B,2008ApJ...689.1327S},
which consider an initial spherical contracting state; and Core Accretion models 
\citep{2007ApJ...655..541M,2008ApJ...683.1104F}, which couple planetary thermal 
evolution to the predicted core mass and thermal structure of a core-accretion 
planet formation model. 

In the following, we only consider the results obtained using the hot start models for
the nearby sample. 
However, the problem of the initial condition and the uncertainties on the stellar ages
are among the main limitations, in case of young stellar samples, not only for our code, 
but also for any kind of study that uses the same kind of approach 
\citep[see e.g.][for a detailed discussion]{2010A&A...509A..52C, 2010A&A...522A...2B}. 
These limitations also apply to the theoretical approach, if the evolutionary models 
are used to evaluate the planet intrinsic luminosity and radii produced by the models, 
as in \cite{2010arXiv1012.5281M, 2011arXiv1102.4146M}.

For the evaluation of the reflected light, we scaled the Jupiter 
value, according with the planet radius (expressed in Jupiter radii), 
semi-major axis, albedo and illuminated fraction of the planet.
This last contribution is computed through a phase dependent term, 
$\Phi(\beta)$, which is given by Eq. \ref{eqn:phi_beta} 
\citep[see][]{2004ApJ...610.1079B}, where $\beta$ is the phase angle 
(angle at companion between star and the observer) and 
$z=r \sin{\left(\nu + \omega \right)}$ is the radial coordinate of the 
radius vector. 
\begin{equation}
\label{eqn:phi_beta}
\Phi(\beta) = \left[\sin \beta + \left(\pi + \beta \right) \cos \beta\right]/\pi 
\end{equation}

The Jupiter/Sun contrast is obtained using Eq. \ref{eqn:ref_light} which gives 
an estimate of the fraction of stellar light captured by a planet, depending 
on the values of the planet radius, semi-major axis and geometrical albedo, being $\Phi(\beta)=1$\
(at opposition).

\begin{equation}
\label{eqn:ref_light} 
(L_{\rm Jup}/L_*)_{\rm Ref}  =  A_{Jup}\frac{R_{\rm Jup}^2}{a_{\rm Jup}^2} = 2.5\times 10^{-9}
\end{equation}

\noindent Where $A_{Jup}=0.35$ is the value of the Jupiter albedo in the the J-Band, 
\citep[see][]{1981JGR....86.8705H}.

Then we end with a final value of the contrast in reflected light given by Eq.\ref{eqn:ref_final}.
\begin{equation}
\label{eqn:ref_final}
(L_{\rm p}/L_*)_{\rm Ref}  =  \left(L_{\rm Jup}/L_*\right)_{\rm Ref}\Phi(\beta)\frac{(R_{\rm p}/R_{\rm Jup})^2}{(a/a_{\rm Jup})^2}
\end{equation}

\noindent As a consequence of Eq.~19, the results of MESS will be sensitive to the choice of $A_{\lambda}$, 
especially for the cold planets, in which the contribution of the reflected light is dominant.
 Following the outcomes of Jupiter observations and theoretical models \citep[See e.g.][]{2004IAUS..202..255B}, 
we decided to uniformly generate the values of the albedo between 0.2 and 0.7. 
The code anyways offers the option to fix the value of the albedo 
to a chosen value, for all the planets in the generation.
A test of the impact of the choice of the albedo value on the redults of the simulations is presented in Sec.~3.4.

\subsubsection{RV and astrometric signal}
 The indirect effects of the presence of the planet, 
such as the semi-amplitude of radial velocity (RV) variations and the astrometric signal
can be inferred, knowing all the orbital characteristics for each planet. 

\subsubsection{Degree of polarization}
 The degree of polarization $\Pi$\ is assumed to be of the form \citep[see e.g.][]{2005A&A...429..713S}:
\begin{equation}
\label{eqn:pol_deg}
\Pi =\Pi_{\rm max}\times(1-\cos^2{\beta)}/(1+\cos^2{\beta})
\end{equation}
where $\Pi_{\rm max}$\ is the maximum polarization value (which is
assumed to be randomly generated between 0.1 and 0.3), and $\beta$ 
is the same as in Eq. \ref{eqn:phi_beta}. Then the contrast due to 
the polarized light of the planets is $\Pi$ times the contribution in 
reflected light evaluated with Eq.~\ref{eqn:ref_final}.

\subsection{Planet population synthesis}
\label{sec:ppsynth}
Depending on the purpose of the analysis, the code can generate the planet population 
in two different ways:
\begin{description}
\item a) {\bf Full population}: the value of $N_{seed}$ sets the spacing of the mass-period grid, 
and for each point on it $N_{gen}$ planets are generated, ending with $N_{seed} \times N_{gen}$ planets 
per each star. The population for each star is saved in an independent file. 
This approach is useful for the statistical analysis of existing data, 
since in this case MESS provides the fraction of detectable planets per 
star, which can be used to derive the global probability of finding a 
planet over the whole target list. This can be then compared with the real results. 

\item b) {\bf Reduced population}: only one orbit is generated for each point in the mass-period grid.
$N_{gen}$ in this case sets the number of planet in a {\it planetary system} associated with each 
star\footnote{note that no consideration on the planet stability is made, and to the purpose
of the analysis each planet is considered separately}.
The final population is then composed by $N_{star} \times N_{seed}$ planets, and all the planets are saved
together in one file. 
Then the predicted detection performances of a given 
instrument can be used, to derive the population of objects that are expected
to be detected around each star, if the whole input sample is observed. 
\end{description}

As an example, we generated a reduced population (assuming 5 planets per star)
of planets around the stars of the nearby sample described in Sec.~\ref{sec:star_pop}.
We choose the semi-empirical approach, and used the {\it typical setup} 
we discussed in Sec.~\ref{sec:mpseed}\footnote{Note that the whole calculation 
of the physical characteristics and observables
described in Sec.~2.2.3 to 2.3.2 can be skipped (with considerable gain in computing 
speed), the code providing in this case only the orbital elements}.

\begin{figure}[tbp]
\center
\includegraphics[width=9.2cm]{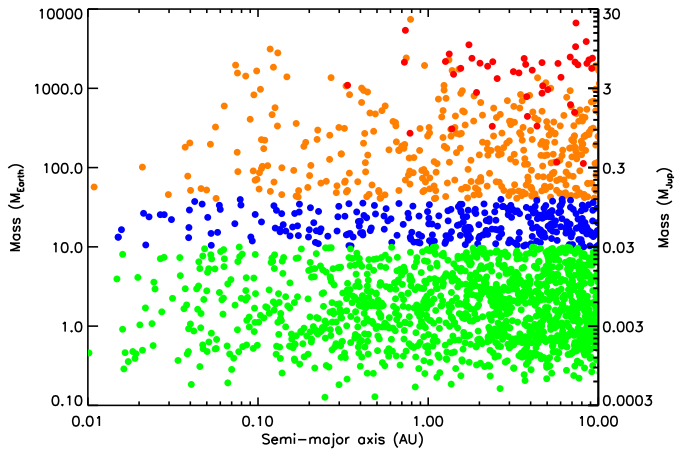}
\caption{\small Mass semi-major axis distribution of the synthetic planets 
in the populations generated by MESS using the semi-empirical approach. 
The different classes of planets (see text)
are plotted using different colours: red/orange for the warm/cold Jupiters, green 
for the Neptune like planets, blue for the rocky planets.}
\label{fig:pop_char_I}
\end{figure}

Fig.~\ref{fig:pop_char_I} shows the position of the planets in the mass vs semi-major axis plane.

The planets are separated into the three classes, using different colours:
\begin{itemize}
\item {\bf Giant (or Jupiter-like)} planets ($M_{\rm planet} > 40 M_{\rm Earth})$.
A distinction between {\it Cold Jupiters} (orange dots) and {\it Warm Jupiters} 
(red dots), as defined in Sec. ~\ref{sec:contrast}, is also made.
\item {\bf Neptune-like} planets ($ 10 M_{\rm Earth} \le M_{\rm planet} \le 
40 M_{\rm Earth}$: green dots) 
\item {\bf Rocky} planets ($M_{\rm planet} < 10 M_{\rm Earth}$: blue dots)
\end{itemize}

\begin{figure}[tbp]
\center
\includegraphics[width=9.2cm]{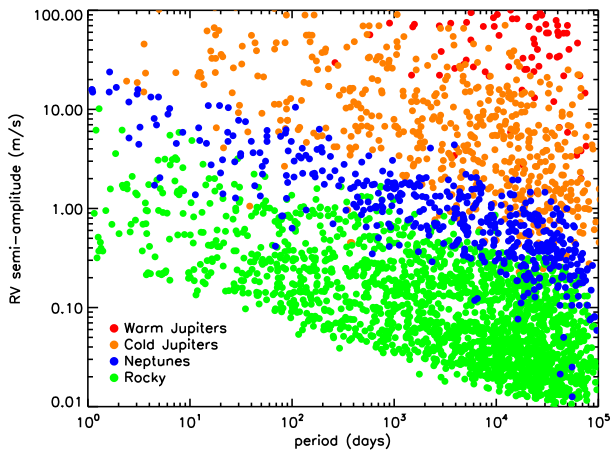}\qquad\qquad
\caption{\small Distribution of radial velocity vs. period of the synthetic planets for the population 
showed in Fig.~\ref{fig:pop_char_I}.}
\label{fig:pop_char_IIIa}
\end{figure}

\begin{figure}[tbp]
\center
\includegraphics[width=9.2cm]{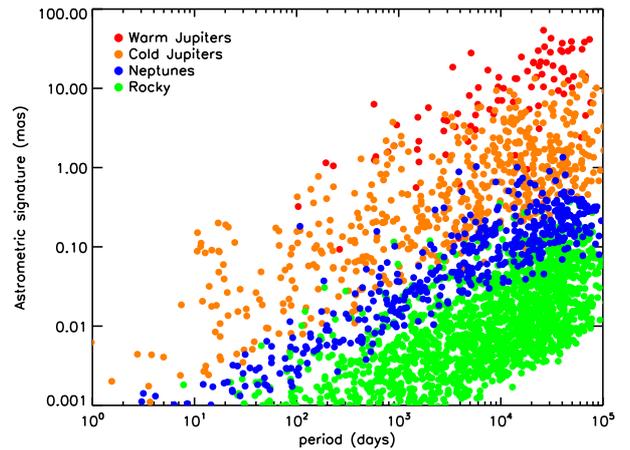}\qquad\qquad
\caption{\small Distribution of the astrometric signal vs. period of the synthetic planets for the population showed in Fig.~\ref{fig:pop_char_I}.}
\label{fig:pop_char_IIIb}
\end{figure}

\begin{figure}[tbp]
\includegraphics[width=9.2cm]{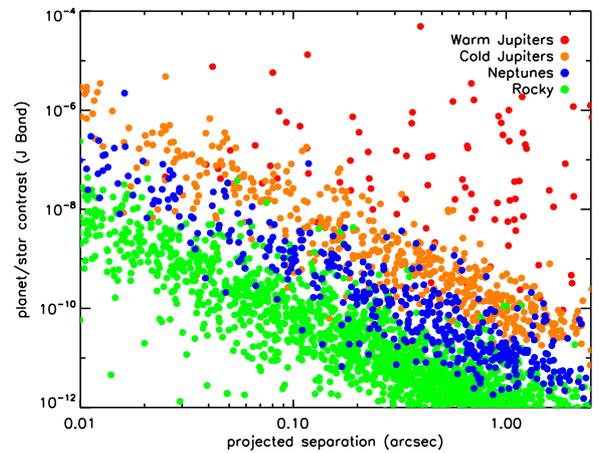}\qquad\qquad
\caption{\small Planet/star contrast vs. projected separation of 
the same planets showed in Fig.~\ref{fig:pop_char_I}.}
\label{fig:pop_char_II}
\end{figure}

The distribution of the observable quantities for the planet showed in Fig.~\ref{fig:pop_char_I} are summarized in Fig.~\ref{fig:pop_char_IIIa}, \ref{fig:pop_char_IIIb}  and~\ref{fig:pop_char_II}.

\subsection{Instrument detection performances}
\label{sec:id_perf}
The last step is the comparison of the observables of the
generated synthetic planets with the detection limits of different
observing techniques, with the possibility to actually combine them. 
It is important, especially in case of comparative studies, 
and since the MESS does not produce the detection limits, to make sure that
the detection performances that are fed as input to the code have been estimated by correctly
taking into account each instrumental biases, specific to each
technique, and the stellar characteristics.
In the
context of the MESS applications, the code has been extensively used
considering two possible inputs for the detection performances:

\begin{itemize}
\item The {\it 1D mode}, which selects the detectable planets using a 
threshold or a curve giving the lower detection limits (RV, astrometric
precisions or contrast performances) as a function of the period, the
semi-major axis, the angular separations etc., defined by the instrumental capabilities 
\item The {\it 2D mode}, which is especially built for the analysis of the 
performances of the Deep Imaging instruments. This mode takes advantage from 
the knowledge of all the orbital elements of the planets, to place them on 
a two dimensional detection map. This mode allows using all the spatial 
information stored in the images. 
Using the whole 2D map not only allows to take into account possible peculiar characteristics of the 
circumstellar environment, such as the presence of disks, but also prevent to under/overestimate the contrast curve 
depending on the method chosen for the extraction itself \citep[see][]{2010A&A...522A...2B}. 
\end{itemize}

\section{Applications}
\label{sec:op_modes}

Once the synthetic population of planets has been created, the next 
step is to compare the characteristics of the generated planets with 
the detection limits appropriate for the instrument under consideration. 

MESS offers three different operation modes (OM), depending on which kind 
of analysis is needed.

\begin{enumerate}
\item The {\bf Hybrid Mode (MESS\_HM)} which is the most flexible one, and
an be used to probe the physical and orbital
properties of a putative companion around one given system based on
the combination of different techniques, a priori information on the
possible orbit given the presence of other planets or circumstellar
disk.
\item The {\bf Statistical Analysis Mode (MESS\_SAM)}, which is built for the 
analysis of  real data and uses the {\it full population} defined in 
Sect.~\ref{sec:ppsynth}. It enables to test different set of planet
populations or constrain the maximum occurrence of planets for a given
population that would be consistent with the results of detection
and/or null-detection of a complete survey of a large target sample.
\item The {\bf Predictive Mode (MESS\_PM)}, which starts from the {\it reduced 
population} (see Sect.~\ref{sec:ppsynth}), and given the predicted 
performances of a planned instruments, can be used to select the most 
suitable targets given the science goals of the instrument itself, to test the results of different observing strategies and finally to foresee possible synergies with other instruments.
\end{enumerate}

\subsection{Single object characterization}
\label{sec:res_hy}
The first and more versatile MESS mode is the so-called Hybrid mode.
This mode can be used for the study of particularly interesting targets, or to test specific hypothesis. 
It allows for example to take into account all available informations about the orbit of a planet already discovered around the target, in order to put constraints on the planet generation.
A preliminary version of this mode has been used to put constraint on the presence of a planetary companion embedded
in the disk surrounding the T-Tauri star LkCa15 \citep[see][]{2010A&A...522A...2B}.

We present here an analogous analysis made for TWA 11. 
This star has been found to be surrounded by a debris disk by \cite{2009AJ....137...53S}. 
Using STIS, \cite{2009AJ....137...53S} provided a full characterization of the disk geometry, and suggested a possible unseen companion 
responsible for some of the observed properties. 
We then decided to use MESS\_HM to verify which kind of constraints can be put using the VLT-NACO observations of this star.

A pixel-to-pixel 2D noise map was estimated from the reduced NACO images, 
using a sliding box of $5\times5$ pixels over the whole FoV.
We then considered a 6~$\sigma$ threshold to build the final detection limit maps to be used 
for the statistical analysis.
These maps were also converted in terms of minimum mass map 
using the evolutionary model predictions at the age of the system.
Fig.~7 shows an example of the resulting sensitivity map\footnote{Note that the decreasing values of the non-detection probability at separations lower than 30-50 AU are due to systematic errors. In fact the detection limit drops to unrealistic low values really close to the star}.

\begin{figure}[tbp]
\label{fig:massmap}
\includegraphics[width=9.5cm]{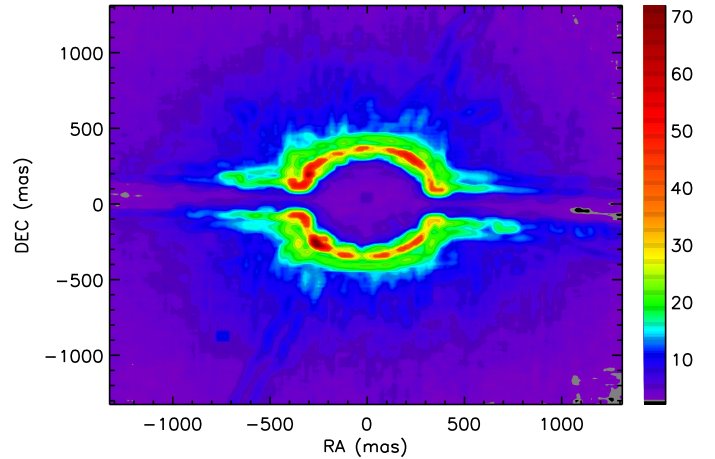}
\caption{2D map giving the values of the minimum mass of detectable companions ($6 \sigma$) as a function of projected separation, around TWA 11}
\end{figure}

We considered only circular orbits coplanar with the disk, with an inclination and a longitude of the ascending node fixed by the disk properties reported in \cite{2009AJ....137...53S}: $i_{Disk}=75.88\pm0.16$, $\Omega = PA \pm 90 = \left(27.1\pm90\right)$.
 \cite{2009AJ....137...53S} also estimated the inner and outer boundary of the disk to be 0.515 and 2.114 arc secs respectively, 
corresponding to 37 and 154 AU at the target distance \citep[72.8 pc, see ][]{2007A&A...474..653V}.
TWA11 is also known to have a stellar companion at $\rho=7.7''$ \citep{1995ApJ...445..451J}. 
As pointed out by \cite{2009AJ....137...53S}, the value of the outer boundary of the disk is consistent with the presence of the companion. 
Using Eq.~\ref{eqn:acrit_cs} we in fact obtained a value for the critical semi-major axis for the planet stability ($a_{crit}$) of about 165 AU. 
Taking into account these constraints, we set the range of explored semi-major axes to 35-160 AU.
The results of our simulations, in terms of non detection probability maps as a function of the companion mass and semi-major axes, are shown in Fig.8. 
The disk boundaries are also shown, as reported by \cite{2009AJ....137...53S}.

Is it clear that with the NACO images we are not able to put strong constraint on 
planetary-mass objects, but surely low-star companions and brown dwarfs more massive than $30 M_{Jup}$  can be excluded at $a > 35 AU$ and $20 M_{Jup}$ ones for $a > 100 AU$.

\begin{figure}[tbp]
\label{fig:TWA11_hyb}
\center
\includegraphics[width=9.2cm]{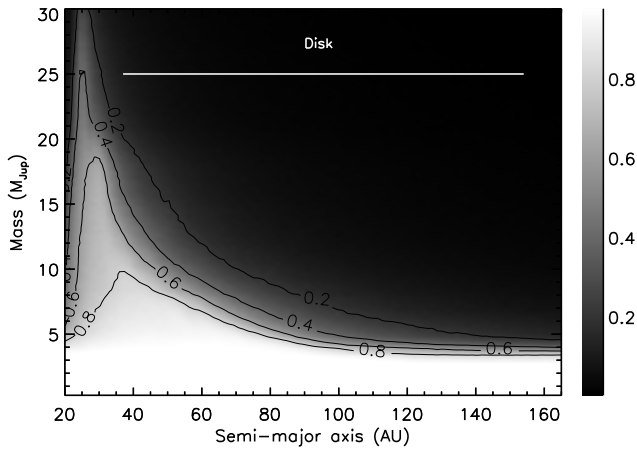}
\caption{Non detection probability map of a faint companion around TWA11 as a function of its mass and semi-major axis, 
in the case of a circular orbit. Inclination and longitude of the ascending node have been fixed using the disk properties 
 $i_{Disk}=75.88\pm0.16$, $\Omega = PA \pm 90 = \left(27.1\pm90\right)$.}
\end{figure}

\subsection{Statistical Analysis of a Survey}
\label{sec:sam}

\subsubsection{Testing the planet population assumption}
\label{sec:sam_se}
The MESS\_SAM operation mode allows to test the consistency of various sets of (mass, 
eccentricity, semi-major axes) parametric distributions of a planet 
population with observational data. Given the detection performances of 
a survey, the frequency of detected simulated planets (over the complete 
sample) enables derivation of the probability of non-detection of a given 
planet population associated with a normalized distribution set. Then
the comparison with the survey results tests directly the disagreement
with observations at an appropriate level of confidence.

As an example of the use of SAM@MES, statistical analysis mode, we present the analysis of a small sample of 
 young neighbourhood stars, part of the TWA association, and observed with NACO@VLT.
These stars are part of a bigger sample for which the observations and statistical analysis, done with a preliminary version of MESS, have been presented by \cite{2010A&A...509A..52C}.
The characteristics of the stars in the sample are listed in Tab.1.  
We present a new analysis of these target, done with the 2D module of MESS\_SAM. 

\begin{table*}[!t]
\caption{Sample of TWA stars considered in our analysis. In addition
to name, coordinates, galactic latitude (\textit{b}), spectral type,
distance, $V$ and $K$ photometry, the observing mode direct imaging
(DI) or coronagraphy (COR) and the status of the primary (single, binary
Bin, triple) are also listed}             
\label{tab:sample_prop_1}      
\centering          
\begin{tabular*}{\textwidth}{@{\excs}lllllllllll}     
\hline\hline       
Name            & $\alpha$        &  $\delta$       &    SpT   & Mass  & \textit{d}  &  Age         & V     & K      &  Mode    &   Notes\\
                & [J2000]         & [J2000]         &          & M$_\odot$   & (pc)        &  (Myr)       & (mag) & (mag)  &          &         \\
\hline  
\hline
\object{ TWA22}           & 10 17 26.9  & -53 54 28   & M5  & 0.15    &  18    &   8  &  13.2  &   7.69   & DI/CI, S27, Ks & Bin ($\rho=0.1''$)   \\
\object{ TWA14}           & 11 13 26.3  & -45 23 43   & M0  & 0.55    &  63    &   8  &  13.8  &   8.50   & DI/CI, S27, Ks &      \\
\object{ TWA12}           & 11 21 05.6  & -38 45 16   & M2  & 0.30    &  32    &   8  &  13.6  &   8.05   & DI/CI, S27, Ks &      \\
\object{ TWA19}           & 11 47 24.6  & -49 53 03   & G5  & 1.50    & 104    &   8  &  9.1   &   7.51   & DI/CI, S13, H  &      \\
\object{ TWA23}           & 12 07 27.4  & -32 47 0    & M1  & 0.40    &  37    &   8  &  12.7  &   7.75   & DI/CI, S13, H  &            \\
\object{ Twa25}           & 12 15 30.7  & -39 48 42   & M5  & 0.15    &  44    &   8  &  11.4  &   7.31   & DI/CI, S27, Ks &            \\
\object{ TWA11}           & 12 36 01.0  & -39 52 10   & A0  & 2.10    &  67    &   8  &  5.8   &   5.77   & DI/CI, S27, H  &  Bin($\rho=7.7''$), Star with disk          \\
\object{ Twa17}           & 13 20 45.4  & -46 11 38   & K5  & 1.00    & 133    &   8  &  12.6  &   9.01   & DI/CI, S13, H  &      \\
\hline       
 \hline                  

\end{tabular*}
\end{table*}

2D minimum mass maps were obtained with the same method used for the analysis of TWA 11 (see Sec.~\ref{sec:res_hy}) 
for all the stars in the sample. We then used MESS\_SAM to calculate the
detection probability ($P_{\rm{D}}$) of companions of various masses
and orbital parameters (semi-major axis $a$, eccentricities $e$,
inclination $i$, longitude of the ascending node $\Omega$, longitude
of periastron $\omega$ and time of periastron passage
$T_p$). 
We used the empirical approach, generating a full population of 10.000 planets for each target, 
with a mass range spanning between 0.3 and 30 $M_{Jup}$ and a cut-off in semi-major axis of 100 AU. 

Each simulated companion was placed on the 2D minimum mass map
according to its position on the projected orbit to test its detectability,
comparing its mass with the minimum value achievable at the same position in the FoV.

Only circumbinary planets were considered around TWA 22, adopting the total mass of the system as $M_{Star}$. 
In fact the binary being so close \citep[$\rho =0.1''$ see][]{2009A&A...506..799B}
leads to a value of the critical semi-major axis 
for circumstellar planets $a_{CS}$ of only 0.456  AU and of  8.395 AU for the circumbinary ones.

Two sets of indices for the power-law distribution were tested:
\begin{enumerate}
\item The ones derived by \cite{2008PASP..120..531C} (CM08): $\alpha=-1.31$, $\beta=-0.74$
\item The ones derived by \cite{2003ApJ...598.1350L} (LW03): $\alpha=-1.81$, $\beta=-0.30$
\end{enumerate}

Finally, fixing $\alpha$ and $\beta$ to the CM08 values, we also introduced different values for the scaling of the planetary mass with the primary mass. 

The results of all these simulations are summarized in Fig.~9.

\subsubsection{Estimate of the frequency of giant planets}

A second more general use is to constrain the
exoplanet fraction $f$\ within the physical separation and mass probed
by a survey, in the case of null or positive detections. Contrary to
what was assumed before, $f$\ becomes an output of the simulation, which
actually depends on the assumed (mass, period, eccentricity)
distributions of the giant planet population.  This statistical
analysis aims at determining $f$, within a confidence range, as a
function of mass and semi-major axis, given a set of individual
detection probabilities $p_j$ directly linked to the detection limits
of each star observed during the survey and to the considered giant
planet distributions.

The probability of planet detection for a survey of $N$ stars can
in fact be described by a binomial distribution, given a success probability
$fp_j$, with $f$\ being the fraction of stars with planets, and $p_j$\ the
individual detection probabilities of detecting a planet if present
around the star $j$. Each individual $p_j$\ can be replaced 
by $\langle p_j \rangle$, the
mean survey detection probability of detecting a planet if
present. Finally, assuming that the number of expected detected
planets is small compared to the number of stars observed ($f \langle
p_j \rangle << 1$), the binomial distribution can be approximated by a
Poisson distribution to derive a simple analytical solution for the
exoplanet fraction upper limit $f_{\rm{max}}$ for a given level of
confidence $\rm{CL}$:

\begin{equation}
\label{eqn:f_max}
\hspace{2.6cm}f_{\rm{max}} = \frac{-\ln{(1-\rm{CL}})}{N \langle p_j \rangle}
\end{equation}

Fig.~10 shows the results obtained applying this module at the sample of stars listed in Tab.~1.

Although the significance of our results is not really high, given the small size of the sample, they are still in agreement with the results of the whole analysis presented by \cite{2010A&A...509A..52C} and with the results of the other deep imaging surveys \citep[see e.g.][]{2010ApJ...717..878N,2007ApJ...670.1367L}.\footnote{An extensive analysis, with MESS\_SAM, of the results of the major deep imaging surveys published in the last decade is ongoing, and will be presented in a forthcoming dedicated paper.}

\begin{figure}[tbp]
\label{fig:SAMprob}
\center
\includegraphics[width=9.5cm]{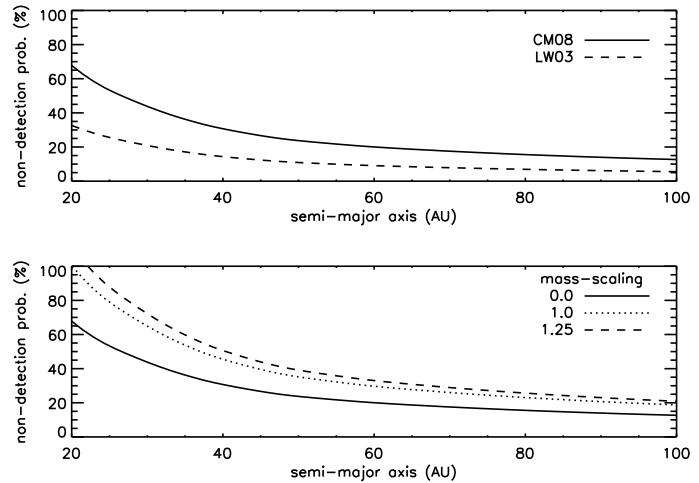}
\caption{Non-detection probability for the stars listed in Tab.~1, based on various sets of period and mass distribution. Mass and period distribution are extrapolated and normalized from RV studies. {\it Top:} Variation of the non detection probability using two different sets of power-law distributions (see text).  {\it Bottom:} variation of the non-detection probability fixing $\alpha = -1.31$ and $\beta = -0.74$ \citep{2008PASP..120..531C} and different scaling the mass of the planet with the primary mass.}
\end{figure}

\begin{figure}[tbp]
\label{fig:gp_fraction}
\center
\includegraphics[width=9.2cm]{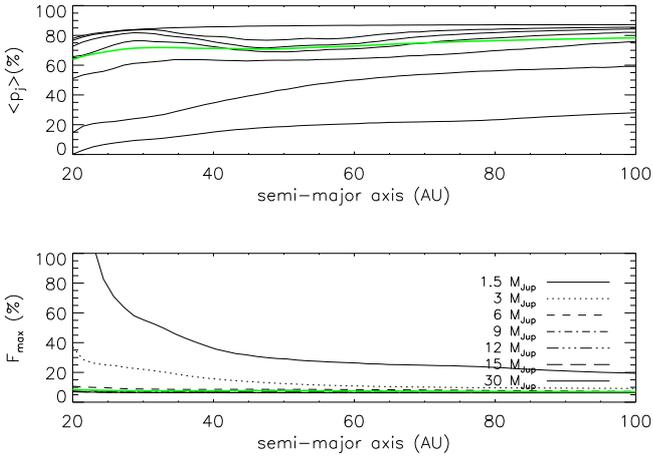}
\caption{{\it Top:} Survey mean detection probability derived as a function of the semi-major axis, assuming parametric mass and period distributions derived by \cite{2008PASP..120..531C}. The results are reported for individual masses: 1.5,3,6,9,12,15,30 $M_{Jup}$. 
The integrated probability for the planetary mass regime is shown with the thick green line. {\it Bottom:} Planet fraction upper limit derived as a function of semi-major axis, given the same mass and period distributions.}
\end{figure}

\subsubsection{Theoretical approach}
\label{sec:sam_te}
The MESS\_SAM can also be used to test the predictions of specific planet formation theories. An extensive use of this OM has been made to analyze a sample of massive stars (B-type and early A-type) observed with NIRI, to test the applicability of planet formation by disk instability in those systems. Starting from a uniform mass versus semi-major axis grid with a sampling of 5 AU~in semi-major axis and $1~M_{\rm jup}$ in mass, $10^4$ orbits were generated for each grid point. Models of disc instability \citep[and Klahr et al., in prep.]{1997ApJ...486..372B,2011arXiv1102.4146M} were then used to provide boundaries in the mass versus semi-major axis space, within which sub-stellar companions can form by this mechanism. These boundaries were dependent on the stellar properties, and so appropriate values should be used for each target in the sample. The planets falling within the allowed range were subsequently evaluated against the 1D detection limits from the high-contrast images of the survey. In this way, by testing a range of planet distributions within the set boundaries, meaningful limits could be placed on the frequency of planet and brown dwarf formation by disk instability in massive disks. The full analysis is presented in detail in the survey paper \citep{2011ApJ...736...89J}. An example of a detection probability map in mass versus semi-major axis space is shown in Fig.~11.

\begin{figure}[tbp]
\label{fig:ari}
\center
\includegraphics[width=9.2cm]{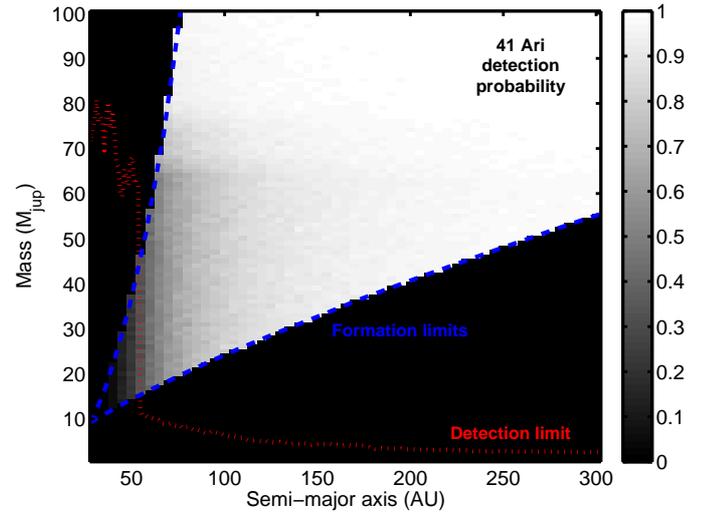}
\caption{Non detection probability map, for 41 Ari (HIP 13209). }
\end{figure}

\subsection{Predictive mode}
\label{sec:pred_om}

\noindent Beside the analysis of the real data, 
MESS can be also used to predict the output of forthcoming searches, 
the goal being to provide information about the capabilities of future planet search instruments.
With this mode, the flexibility of the code reaches its maximum, providing a wide range of possible applications.

Once the synthetic planet population has been created, and assuming the characteristics of a given instrument,
MESS\_PM allows predicting the number of detections expected from a 
future facility.
This provides informations on :

\begin{enumerate}
\item the expected frequency of planets
\item the properties of these objects 
\item the kind of constraints that their observation can put on the planet 
formation theories.
\end{enumerate} 
Furthermore, it also allows to test different instrumental configurations 
and observational strategies that can be adopted, 
thus providing a tool to tune the instrument characteristics, in 
order to fulfil the requirements needed to access a certain domain in 
the parameter space, and reach the proposed science goals.

\subsubsection{Comparison of future direct imaging instrument capabilities}

As an example of the application of MESS\_PM, we report the 
results of a comparison of the capability of a set of instruments for the direct imaging of exoplanets, 
planned for the next decade, which are briefly described in Tab.~2.

\begin{table}[tbp]
\caption{\footnotesize Instruments for direct imaging of exoplanets considered in our analysis. 
References: {\bf B10} \cite{2010ASPC..430..231B};
{\bf G05} \cite{2005SPIE.5905..185G};
{\bf G07} \cite{2007AAS...21113402G};
{\bf K10} \cite{2010SPIE.7735E..81K};
{\bf R10} \cite{2010AAS...21543904R};
{\bf S10} \cite{2010SPIE.7736E.115S};
{\bf T10} \cite{2010AAS...21642501T}
}
\begin{tabular}{lccccl}
\hline
\hline
Instrument  & Contrast   & Wavel.  & IWA	&  Year & Ref. \\
            &            & ($\mu$m)    & ('') & \\
\hline
\hline
\multicolumn{5}{c}{8~m ground-based telescopes}\\
\hline
VLT-SPHERE & 10$^{-7}$ & 0.9-1.7 & 0.08    & 2011 & B10\\
Gemini-GPI &           &         &         &     &  G07  \\
\hline
\hline
\multicolumn{5}{c}{JWST}\\
\hline
NIRCAM     & 10$^{-5}$ & 2.1-4.6 & 0.30    & 2014 & G05\\
MIRI       & 10$^{-4}$ &  5-25   & 0.35    &      & R10    \\
\hline
\hline
\multicolumn{5}{c}{1.5 m Space Coronagraphs}\\
\hline
	       & $10^{-9}-10^{-10}$ & 0.3-1.3 & 0.08 & ? & T10\\
\hline
\hline
\multicolumn{5}{c}{ELT's class instruments}\\
\hline
E-ELT-EPICS & $10^{-8}-10^{-9}$& 0.9-1.7 & 0.03 & $>2018$ & K10\\
E-ELT-METIS & $10^{-5}$& 2.5-20 & 0.08 & & S10\\
\hline
\hline
\end{tabular}
\label{tab:instr}

\end{table}

Since the purpose of the presented analysis is purely illustrative, 
we adopted for each instrument an averaged detectability relation, taken from the reference indicated in Tab.~2, then using only the 1D approach. 
The sample of stars used is the one described in Sec.~\ref{sec:mpseed} and whose properties are summarized in 
Fig.~\ref{fig:sample}.
This sample was originally selected as a preliminary sample for the planet search survey to be done 
with SPHERE, the next generation planet finder of VLT \citep{2008SPIE.7014E..41B}, and it's therefore optimized for this 
kind of instruments, which possibly introduces some biases against some of the other instrument analyzed. 
The analysis was made using the reduced population, assuming five planet per star. 

The results of the analysis, showed in Fig.~12 and also summarized in Tab.~3, foresee that enormous progress that can be expected in the next decade.
The available measurements are already giving us indirect information on far away planets around young stars, but passing through the intermediate step of next generation image and finally with the advantage of ELT instruments we will have a wide view on planetary systems at different stages of their evolution.

\begin{figure*}[tbp]
\center
\label{fig:PMnearby}
\includegraphics[width=18cm]{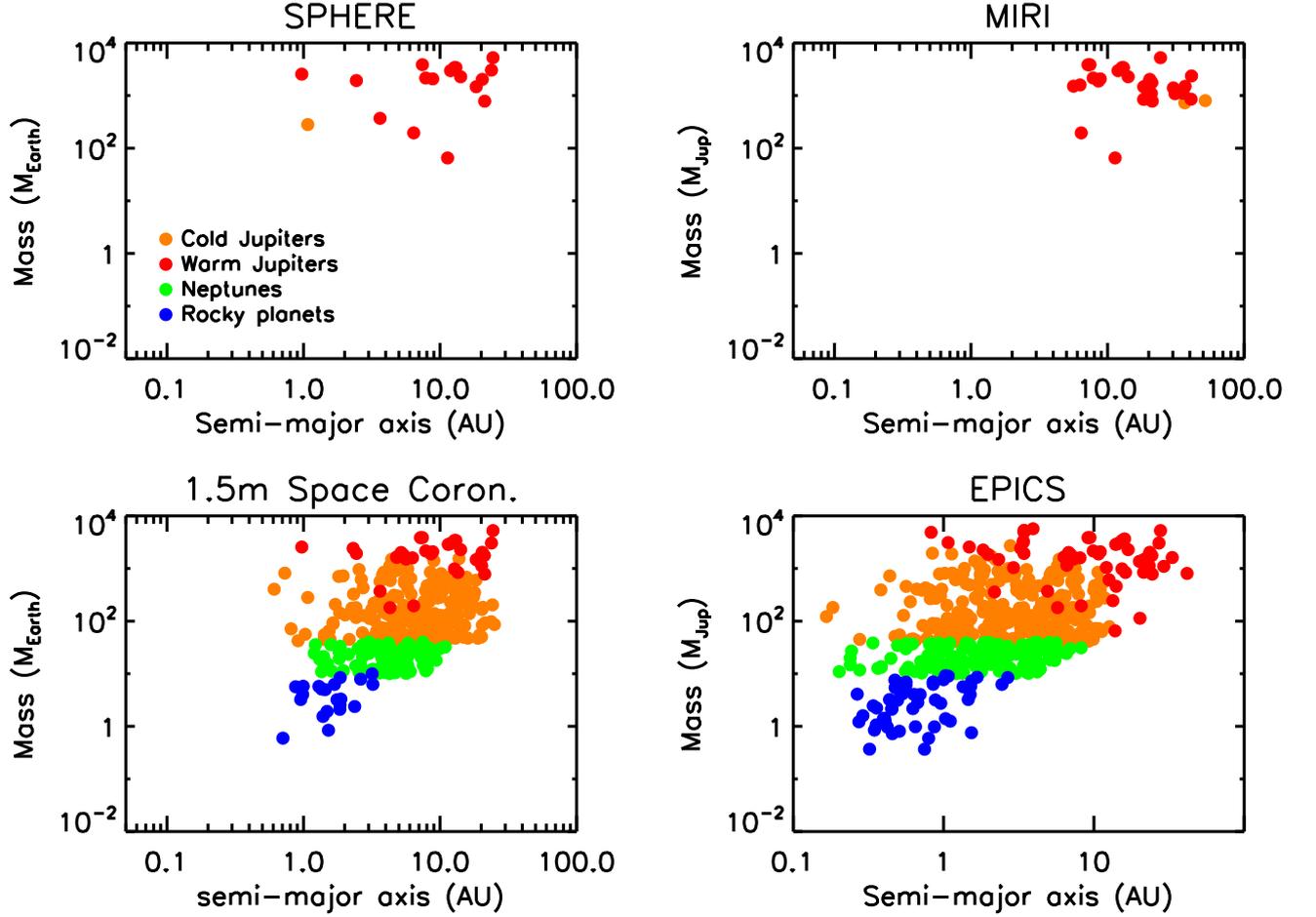}
\caption{Planets expected to be discovered by SPHERE (representative of planet finders on 8m
class ground-based telescopes), JWST-MIRI, 1.5m Space Coronagraphs, and EPICS/E-ELT
(representative of 30-40m class telescopes) in the mass vs separation plane. Different
colours are used for warm giant (orange), cold giant (red), Neptune-like (green), and rocky
planets (blue) respectively.}
\end{figure*}

\begin{table}[htbp]
\caption{\footnotesize Summary of expected detections from imagers in the next decade}
\label{tab:res_PM}
\begin{tabular}{lcccccc}
\hline
\hline
Instrument       & Year	& Young  & Old      & Nept. & Rocky   \\
                 &      & Giants & Giants   &   &  \\
\hline
Gr. based 8m	 & 2011	& tens	      & few       &	     &     \\	
JWST	         & 2014  & tens	      & few       &	     &	   \\
1.5m Space Coro. & ?  	& tens	      & tens      & tens & few \\
ELT's            & $>2018$ & hundr. & hundr. & tens & few \\
\hline
\end{tabular}
\end{table}

\subsubsection{Foreseeing the synergies between different techniques}

\noindent Once the RV and astrometric modules will be completed, 
MESS\_PM  will provide an estimate of both the direct and indirect signatures 
of the presence of the planets, and thus be used to compare the outcomes of imaging with dynamical methods. 
These are interesting, because the latter allow determining the planet masses, thus eliminating the degeneracy with age,
which is currently one of the major problems affecting direct detections. 
Moreover, possible synergies between different discovery methods are becoming more and more likely, 
ELT's instruments representing the ideal link between direct and indirect detections, 
covering both young, nearby systems discovered by next generation 
imagers and also meant to provide the first images of planets already detected by RV.

Fig.~13 and ~14 summarize the results of the preliminary version of the RV and astrometric modules of 
MESS\_PM. The planets showed are the same as in the lower right panel of Fig. 12. 

If confirmed, these results would suggest that the discovery space for EPICS at E-ELT overlaps well with those from radial velocity (RV)
instruments (HARPS at ESO 3.6m telescope, ESPRESSO at VLT, and especially CODEX at E-ELT) as well as with that of GAIA
\citep[][]{2008A&A...482..699C}. 

The RV module being still under test, and without having enough data to perform a consistent and accurate analysis 
of the performances and comparison between the instruments under scrutiny, this analysis is not meant 
to tell which instrument is going to provide the highest number of detection, 
but just at showing the potential of the further versions of the code.

\begin{figure}[!ht]
\begin{center}
\includegraphics[width=9.2cm]{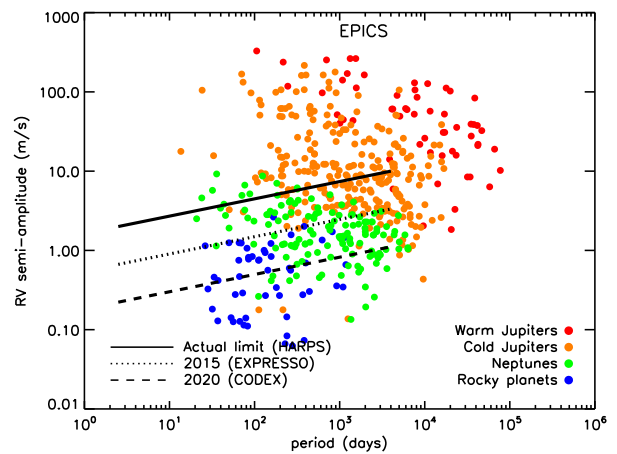}
\end{center}
\caption{Planets expected to be detected by EPICS (nearby sample) in the RV signal vs.
period plane, compared with detection limits for RV instruments (HARPS, ESPRESSO and
CODEX). The colour code is the same as in Fig.~1.}
\label{fig:PM_rv}
\end{figure}

\begin{figure}[tbp]
\label{fig:PM_as}
\center
\includegraphics[width=9.2cm]{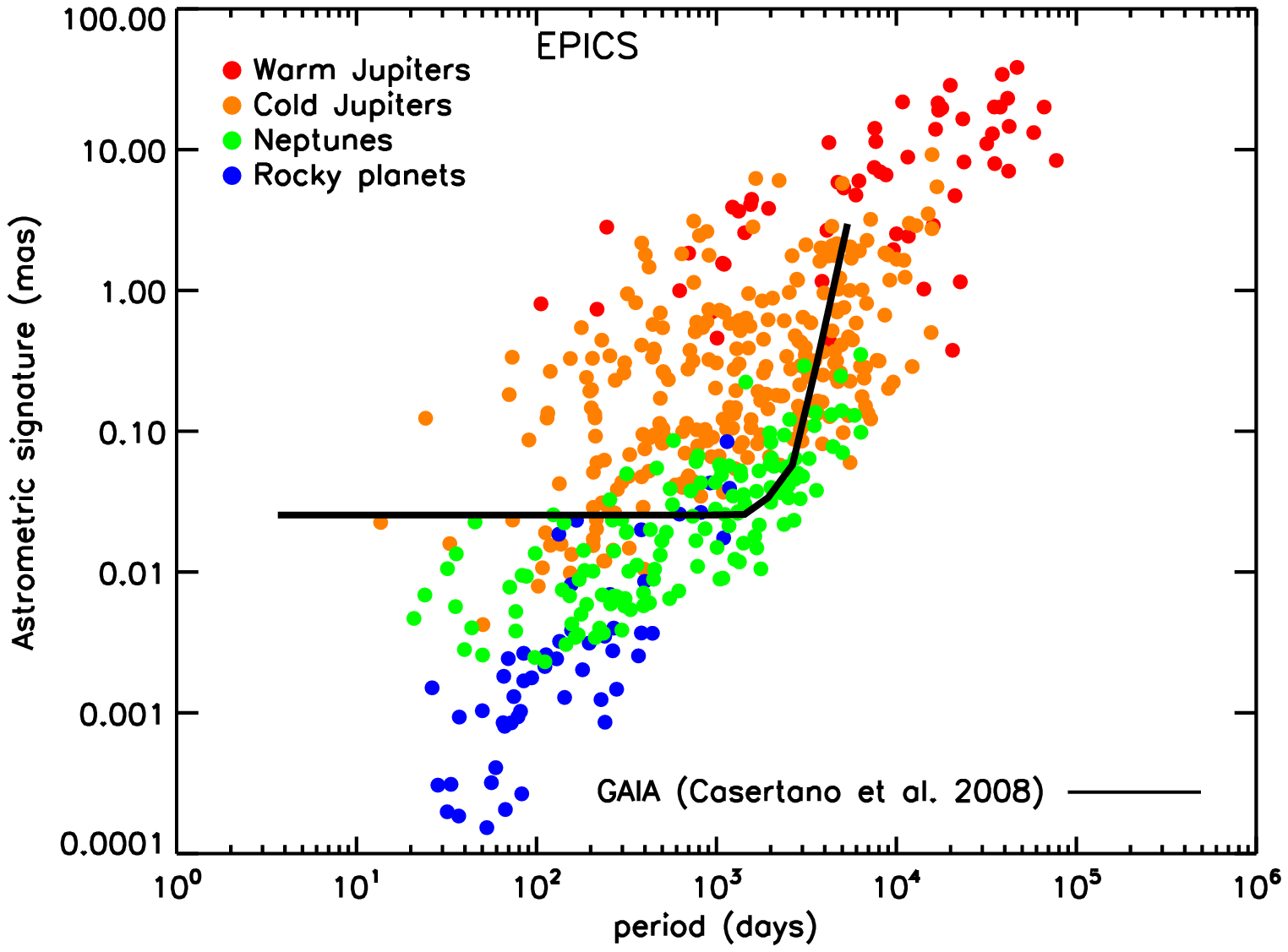}
\caption{Planets expected to be detected by EPICS (nearby sample) in the astrometric signal
vs. period plane, compared with detection limits for astrometric satellites GAIA. Colour code is the same as in Fig.~1.}
\end{figure}

\subsection{Testing the influence of the physical inputs}
In this last section we present the results of some tests which goal was to show how MESS  
can be used to investigate the influence of the various physical parameters considered as inputs 
for the planet generation. 
In particular we focused on the eccentricity distribution and on the value of the planetary albedo. 

\subsubsection*{Eccentricity distribution}
Direct Imaging surveys are, by definition, mostly sensitive to planets in wide orbits. Also, planets on highly eccentric orbits could also be preferred targets, since they are more likely to be found farther out with respect to planets on a circular orbit with the same semi-major axis. 
This could led to a bias towards high eccentricity planets in our results. 
As mentioned in Sec.~2.2, the eccentricity distribution of the planets generated by MESS is uniform, and cut at $e=0.6$. 

\begin{figure}[!ht]
\begin{center}
\includegraphics[width=9.2cm]{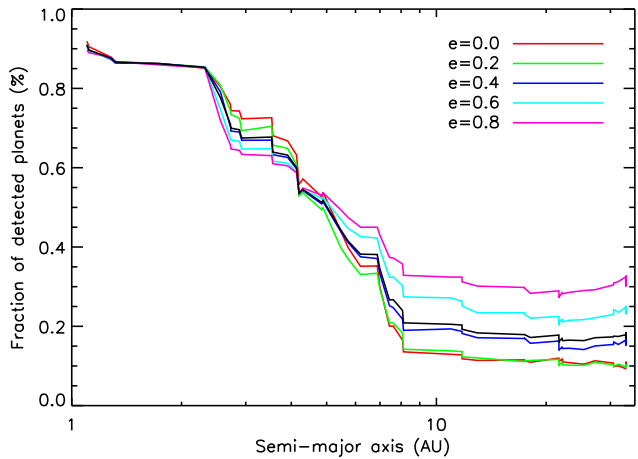}
\end{center}
\caption{Fraction of detected planet as a function of the semi-major axis 
        value (AU) for different values of the planet eccentricity.
        The black solid line shows the results obtained with the 
        {\it standard setup}.}
\label{fig:ecc_test}
\end{figure}

As a further check of the impact of the eccentricity on the DI results,
we repeated the analysis of the TWA sample done in Sec.~3.2, by fixing the eccentricity for all the generated planets to a given value. 
The results are showed in Fig.~15. 
The black solid line shows the results of the {\it standard setup}, thus with the uniform eccentricity distribution cut at $e=0.6$. The red, green, blue, purple and light blue lines show the outcomes of the simulations done by fixing $e=0$, $e=0.2$, $e=0.4$, $e=0.6$ and $e=0.8$, respectively. 
As expected, the higher eccentricity values can lead to an higher fraction of detected planets, for a given semi-major axis value. 

This simple exercise shows not only that the {\it standard setup} of the MESS does not introduce any systematic bias towards high eccentricity, but also that the code allows us to easily take this kind of biases into account, if they are proven to be real, by changing the simulation parameters. 

As a final remark, it has to be said that the effect of the eccentricity is important only in the case of {\it warm} planets, as the ones that could be found around our TWA targets. As the age of the stars increases, the reflected light contribution to the planet contrast becomes more and more important, thus counterbalancing the effect of the eccentricity. 

 \subsubsection*{Albedo distribution}

As mentioned in Sec.~2.3, the Albedo of the planets in the synthetic population is randomly generated between 0.2 and 0.7. 
Especially for the {\it cold} planets, the value of the albedo can be a critical parameter for the planet detection. 
We then decided to perform a check to see how big is the impact on the simulation results.
With an approach similar to the one used to test the eccentricity effect (see Sec.~3.3.3), we performed different sets of simulation, $A_{\lambda}$ being the only free parameter. 
We used an hypothetical G2V star ($J=$, $age=4.5~Gyrs$) at 20 parsecs as target, and the detection limits of EPICS (see Tab.~2).

\begin{figure}[!ht]
\begin{center}
\includegraphics[width=9.2cm]{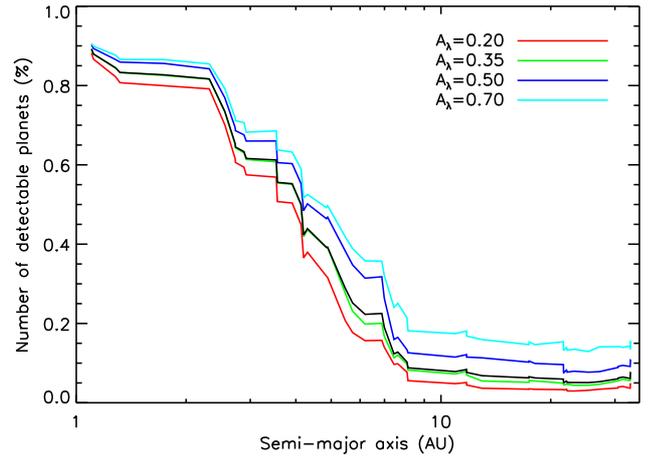}
\end{center}
\caption{Fraction of detected planet as a function of the semi-major axis 
        value (AU) for different values of the planet albedo ($A_{\lambda}$). 
        The black solid line shows the results obtained with the 
        {\it standard setup}.}
\label{fig:alb_test}
\end{figure}

Fig.~16 shows the results of the {\it standard setup} ($A_{\lambda}$ randomly generated between 0.2 and 0.7, black solid line), together with the ones obtained by fixing $A_{\lambda}$ to $0.2$, $0.35$ (the Jupiter value, see Sec.~2.3.1), $0.5$ and $0.7$ (red, green, blue and purple line, respectively). 

As expected, the fraction of detectable planets is higher for higher albedo, all the other parameters being the same. 

This test confirms that the use of an uniform distribution allow us not to favour planets with high albedo, the results being similar to the one obtained by using the Jupiter value.

\section{Summary and conclusions}
\label{sec:conclusions}
In this paper, we presented MESS (Multi-purpose Exoplanet Simulation System), 
a Monte Carlo tool for the statistical analysis and prediction of survey 
results for exoplanets. 

Our aim was to build an extremely versatile code, that could be used to test 
the outcomes of any instrument/technique for the detection of planets.
We consider several assumptions on:
\begin{itemize}
\item The star population, and how to take into account the properties of each star 
and their effect on either the characteristics of the planets or the instrument 
capabilities. The binarity aspects is also included to take into account the 
possible effects of a stellar companion to the planet formation.
\item The planet population, providing the complete set of orbital elements 
and a large number of physical parameters of the planets (radius, temperature, luminosity, etc.), 
either generated using the information coming from the analysis of the planets confirmed up to now (semi-empirical approach) or using the results of the planet formation theories (theoretical approach).
\item The predicted observables (luminosity and polarimetric contrast, RV semi-amplitude, astrometric signal)
\item The synthesis of a planet population, that can be easily adapted to the purpose of the investigation
\item The final comparison with the detection
limits, with the possibility to combine the informations coming from different observing techniques, to select a sub-sample of {\it detectable} planets whose characteristics can then be investigated. 
\end{itemize}

\noindent The code is such that each and every one of these assumptions can be released and/or changed. 
This not only provides a tool which is independent from the models 
(e.g. the planet formation theory chosen if the theoretical approach is used, or the evolutionary 
models used to estimate the planet luminosity and radius) but also makes it relevant to test model 
prediction, as well as to constraint the properties of the known planets under different initial conditions.

So far only the Direct Imaging module of the code has been extensively used, 
but the combination of various techniques is under test
and will offer rich perspective for future combined studies of exoplanets.

Three main applications of the MESS code have been shown:
\begin{enumerate}
\item The Hybrid mode, built for the analysis of single objects, is presented in Sec.\ref{sec:res_hy}.
It can be used to probe the
physical and orbital properties of a putative companion around a
given system based on the combination of different techniques, and possibly a priori
information on the orbit given the presence of other planets
or of a circumstellar disk.
\item The SAM mode (Sec.\ref{sec:sam}), optimized for the analysis of a large sample of stars, 
shows its full potential in Sec.~\ref{sec:sam}, by providing a detailed statistical analysis of a sample of stars observed with direct imaging.
Both the agreement of the observations with the observed parameter distributions (Sec.~\ref{sec:sam_se}) and the planet formation theories (Sec.~\ref{sec:sam_te}) are tested, using the semi-empirical and theoretical approach, respectively. 
\item The PM mode finally aims at the prediction of the outcomes of future searches, 
and can be used to tune not only the main instrument parameters, 
but even the observing strategy.
\end{enumerate}

\noindent However, an extensive use of the code requires
a complete knowledge of the instrument under test, of all the error
sources and of the detection capabilities. Then, to really extend the
use of MESS to other facilities one should first properly set all the
needed parameters. As already mentioned before, both the RV and the astrometric part are currently
included in a very simplistic way. A better treatment of the dependence
of the detectability with astrometry from the orbital parameters should be included. 
A rigorous treatment of the stellar jitter evaluation must
be implemented to allow a better comparison between the imaging
and radial velocity capabilities. Especially in the case of E-ELT instruments,
this would allow to better define the synergies between
the various channels, for a more focused observing strategy.

Moreover, a precise measure of the stellar characteristics is also needed, 
in order to minimize the effects that errors on these parameters, such as the age 
of the system or the presence of stellar companions, 
can have on the analysis. 

Finally, the inclusion of an analysis of the planet stability in case of multiple objects is planned, 
together with an extensive use of the theoretical approach, using the outcomes of the most recent Bern models \citep{2010arXiv1012.5281M}.

Each technique performances vary with
the star properties (age, mass, distance...), have different
observables (luminosity, minimum mass, radius...), different observing
strategies. It is therefore extremely important to take this into
account to ring the maximum constrains, first on the properties of
giant planets (physical, orbital parameter space) that will actually
entirely shape the planetary system architecture, then possibly on the
telluric planets. A better characterization of the giant and telluric
planet orbital and physical properties, including their dependency
with the host properties, is critical for a better understanding of
their formation processes as various mechanisms may be at play 
\citep{2009ApJ...695L..53B,2009ApJ...707...79D,2009MNRAS.400.1563S}, but
also of their architecture and dynamical
evolution. At the end, one additional and important issue is to
understand the required physical conditions that will lead to the
formation of telluric planets in habitable zone within planetary
systems shaped by giant planets, and that will possibly lead to the
formation of Life.

\begin{acknowledgements} This work was done as part of M. Bonavita PhD thesis, which was founded
by the Italian Institute for Astrophysics (INAF). 

The authors would like to thank Prof. W. Benz and Dr. Y. Alibert
for providing essential inputs for the development of the 
theoretical approach of the MESS.

We also acknowledge support from the French National 
Research Agency (ANR) through project grant ANR10-BLANC0504-01 \end{acknowledgements}

\bibliographystyle{aa}

\bibliography{ms}

\end{document}